\documentclass[twocolumn, times]{aastex631}
\usepackage{xspace}
\usepackage{amsmath}
\usepackage{amssymb}
\usepackage{bm}
\usepackage{hyperref}

\newcommand{\tess}{\emph{TESS}\xspace}

\providecommand{\bjdtdb}{\ensuremath{\rm {BJD_{TDB}}}}

\providecommand{\teff}{\ensuremath{T_{\rm eff}}}

\providecommand{\vsinistar}{\ensuremath{v\sin I_\star}}

\providecommand{\msun}{\ensuremath{M_\Sun}}
\providecommand{\rsun}{\ensuremath{R_\Sun}}

\providecommand{\rj}{\ensuremath{R_{\rm Jup}}}
\providecommand{\mj}{\ensuremath{M_{\rm Jup}}}

\newcommand{\ms}{\,m\,s$^{-1}$}
\newcommand{\kms}{\,km\,s$^{-1}$}

\newcommand{\tic}{TIC 147660886\xspace}

\newcommand{\toi}{TOI-2005\xspace}
\newcommand{\toib}{TOI-2005 b\xspace}

\newcommand{\pmRA}{$-25.248\pm0.012$}
\newcommand{\pmDEC}{$8.399\pm0.016$}
\newcommand{\parallax}{$2.991\pm0.017$}
\newcommand{\vsiniLSD}{$111\pm1$}
\newcommand{\loggfit}{$3.946\pm0.035$}
\newcommand{\tefffit}{$7130\pm150$}

\newcommand{\mstar}{$1.59^{+0.016}_{-0.017}$}
\newcommand{\rstar}{$2.02^{+0.14}_{-0.16}$}
\newcommand{\lstar}{$9.17^{+0.52}_{-0.99}$}
\newcommand{\age}{$1.6\pm0.1$}
\newcommand{\dist}{328.7$^{+4.6}_{-4.2}$}

\newcommand{\per}{$17.305904_{-0.000020}^{+0.000023}$}

\newcommand{\plrad}{$1.07^{+0.06}_{-0.11}$}
\newcommand{\ecc}{$0.597^{+0.097}_{-0.065}$}
\newcommand{\semimaj}{$0.16\pm0.02$}
\newcommand{\lam}{$4.8^{+2.3}_{-2.5}$}

\shorttitle{TOI-2005}
\shortauthors{Bieryla, et al.}

\begin{document}

\title{TOI-2005b: An Eccentric Warm Jupiter in Spin-Orbit Alignment}

\correspondingauthor{Allyson Bieryla}
\email{abieryla@cfa.harvard.edu}

\newcommand{\CfA}{Center for Astrophysics \textbar \ Harvard \& Smithsonian, 60 Garden Street, Cambridge, MA 02138, USA}
\newcommand{\USQ}{University of Southern Queensland, Centre for Astrophysics, West Street, Toowoomba, QLD 4350 Australia}
\newcommand{\FlatironCCA}{Center for Computational Astrophysics, Flatiron Institute, 162 Fifth Avenue, New York, NY 10010, USA}
\newcommand{\UIUC}{Department of Astronomy, University of Illinois at Urbana-Champaign, Urbana, IL 61801, USA}
\newcommand{\PSUAA}{Department of Astronomy \& Astrophysics, 525 Davey Laboratory, The Pennsylvania State University, University Park, PA, 16802, USA}
\newcommand{\PSUCEHW}{Center for Exoplanets and Habitable Worlds, 525 Davey Laboratory, The Pennsylvania State University, University Park, PA, 16802, USA}
\newcommand{\NOIRLAB}{NSF’s National Optical-Infrared Astronomy Research Laboratory, 950 N. Cherry Avenue, Tucson, AZ 85719, USA}
\newcommand{\MITKavli}{Department of Physics and Kavli Institute for Astrophysics and Space Research, Massachusetts Institute of Technology, Cambridge, MA 02139, USA}
\newcommand{\DTU}{DTU Space,  Technical University of Denmark, Elektrovej 328, DK-2800 Kgs. Lyngby, Denmark}
\newcommand{\CarnegieOBS}{The Observatories of the Carnegie Instution for Science, 813 Santa Barbara Street, Pasadena, CA, 91101, USA}
\newcommand{\CarnegieLCO}{Las Campanas Observatory, Carnegie Institution for Science, Colina El Pino, Casilla 601 La Serena, Chile}
\newcommand{\CarnegieEPL}{Earth and Planets Laboratory, Carnegie Institution for Science, 5241 Broad Branch Road, NW, Washington, DC 20015, USA}

\author[0000-0001-6637-5401]{Allyson Bieryla}
\affiliation{\CfA}
\affiliation{\USQ}

\author[0000-0002-3610-6953]{Jiayin Dong} 
\altaffiliation{Flatiron Research Fellow}
\affil{\FlatironCCA}
\affil{\UIUC}

\author[0000-0002-4891-3517]{George Zhou} 
\affiliation{\USQ}

\author[0000-0003-3773-5142]{Jason D.\ Eastman} 
\affil{\CfA}

\author[0000-0002-4321-4581]{L. C. Mayorga} 
\affiliation{The Johns Hopkins University Applied Physics Laboratory, 11100 Johns Hopkins Rd, Laurel, MD, 20723, USA}

\author[0000-0001-9911-7388]{David W. Latham} 
\affil{\CfA}

\author[0000-0003-0035-8769]{Brad Carter} 
\affil{\USQ}

\author[0000-0003-0918-7484]{Chelsea X. Huang} 
\affil{\USQ} 


\author[0000-0002-8964-8377]{Samuel N. Quinn} 
\affil{\CfA}

\author[0000-0001-6588-9574]{Karen A.\ Collins}  
\affil{\CfA} 


\author[0000-0002-0856-4527]{Lyu Abe} 
\affil{Observatoire de la C\^ote d'Azur, UniCA, Laboratoire Lagrange, CNRS UMR 7293, CS 34229, 06304 Nice cedex 4, France}

\author{Yuri Beletsky} 
\affil{\CarnegieLCO}


\author[0000-0002-9158-7315]{Rafael Brahm} 
\affil{Facultad de Ingenier\'{i}a y Ciencias, Universidad Adolfo Ib\'{a}\~{n}ez, Av. Diagonal las Torres 2640, Pe\~{n}alol\'{e}n, Santiago, Chile}
\affil{Millennium Institute for Astrophysics, Chile}
\affil{Data Observatory Foundation, Chile}

\author[0000-0001-8020-7121]{Knicole D. Col\'{o}n} 
\affiliation{NASA Goddard Space Flight Center, Exoplanets and Stellar Astrophysics Laboratory (Code 667), Greenbelt, MD 20771, USA}


\author[0000-0002-2482-0180]{Zahra~Essack} 
\affiliation{Department of Physics and Astronomy, The University of New Mexico, 210 Yale Blvd NE, Albuquerque, NM 87106, USA}

\author[0000-0002-7188-8428]{Tristan Guillot} 
\affil{Observatoire de la C\^ote d'Azur, UniCA, Laboratoire Lagrange, CNRS UMR 7293, CS 34229, 06304 Nice cedex 4, France}

\author[0000-0002-1493-300X]{Thomas Henning} 
\affil{Max-Planck-Institut f\"{u}r Astronomie, K\"{o}nigstuhl  17, 69117 Heidelberg, Germany}

\author[0000-0002-5945-7975]{Melissa J.\ Hobson} 
\affil{Max-Planck-Institut f\"{u}r Astronomie, K\"{o}nigstuhl  17, 69117 Heidelberg, Germany}
\affil{Millennium Institute for Astrophysics, Chile}

\author[0000-0003-1728-0304]{Keith Horne} 
\affiliation{SUPA Physics and Astronomy, University of St. Andrews, Fife, KY16 9SS Scotland, UK}

\author[0000-0002-4715-9460]{Jon~M.~Jenkins}  
\affiliation{NASA Ames Research Center, Moffett Field, CA 94035, USA}

\author{Mat\'ias I. Jones} 
\affiliation{European Southern Observatory, Alonso de C\'ordova 3107, Vitacura, Casilla,19001, Santiago, Chile}

\author[0000-0002-5389-3944]{Andr\'es Jord\'an}  
\affil{Facultad de Ingenier\'{i}a y Ciencias, Universidad Adolfo Ib\'{a}\~{n}ez, Av. Diagonal las Torres 2640, Pe\~{n}alol\'{e}n, Santiago, Chile}
\affil{Millennium Institute for Astrophysics, Chile}
\affil{Data Observatory Foundation, Chile}

\author{David Osip} 
\affil{\CarnegieLCO}

\author[0000-0003-2058-6662]{George~R.~Ricker}  
\affiliation{\MITKavli}

\author[0000-0001-8812-0565]{Joseph E. Rodriguez} 
\affiliation{Center for Data Intensive and Time Domain Astronomy, Department of Physics and Astronomy, Michigan State University, East Lansing, MI 48824, USA}

\author[0000-0002-7382-0160]{Jack Schulte} 
\affiliation{Center for Data Intensive and Time Domain Astronomy, Department of Physics and Astronomy, Michigan State University, East Lansing, MI 48824, USA}

\author{Richard P. Schwarz} 
\affil{\CfA} 

\author[0000-0002-6892-6948]{Sara~Seager} 
\affiliation{\MITKavli}
\affiliation{Department of Earth, Atmospheric and Planetary Sciences, Massachusetts Institute of Technology, Cambridge, MA 02139, USA}
\affiliation{Department of Aeronautics and Astronautics, MIT, 77 Massachusetts Avenue, Cambridge, MA 02139, USA}


\author[0000-0002-1836-3120]{Avi Shporer}  
\affil{\MITKavli}

\author[0000-0002-3503-3617]{Olga Suarez} 
\affil{Observatoire de la C\^ote d'Azur, UniCA, CNRS, CS 34229, 06304 Nice cedex 4, France}

\author[0000-0001-5603-6895]{Thiam-Guan Tan} 
\affil{Perth Exoplanet Survey Telescope, Perth, Western Australia, Australia}



\author[0000-0002-8219-9505]{Eric B. Ting}  
\affiliation{NASA Ames Research Center, Moffett Field, CA 94035, USA}

\author[0000-0002-5510-8751]{Amaury Triaud} 
\affil{School of Physics \& Astronomy, University of Birmingham, Edgbaston, Birmingham, B15 2TT, UK}

\author{Andrew Vanderburg} 
\affil{\MITKavli} 


\author{Jesus Noel Villase{\~ n}or} 
\affiliation{Department of Physics and Kavli Institute for Astrophysics and Space Research, Massachusetts Institute of Technology, Cambridge, MA 02139, USA}
 
\author[0000-0002-0701-4005]{Noah Vowell} 
\affiliation{Center for Data Intensive and Time Domain Astronomy, Department of Physics and Astronomy, Michigan State University, East Lansing, MI 48824, USA}
\affiliation{\CfA}

\author[0000-0001-8621-6731]{Cristilyn N.\ Watkins}
\affil{\CfA} 

\author[0000-0002-4265-047X]{Joshua~N.~Winn}  
\affiliation{Department of Astrophysical Sciences, Princeton University, 4 Ivy Lane, Princeton, NJ 08544, USA}

\author[0000-0002-0619-7639]{Carl Ziegler} 
\affiliation{Department of Physics, Engineering and Astronomy, Stephen F. Austin State University, 1936 North St, Nacogdoches, TX 75962, USA}

\begin{abstract}

We report the discovery and characterization of \toib, a warm Jupiter on an eccentric ($e\sim0.59$), 17.3-day orbit around a $V_\mathrm{mag}$ = 9.867 rapidly rotating F-star. The object was detected as a candidate by TESS and the planetary nature of \toib was then confirmed via a series of ground-based photometric, spectroscopic, and diffraction-limited imaging observations. The planet was found to reside in a low sky-projected stellar obliquity orbit ($\lambda = $\lam\ degrees) via a transit spectroscopic observation using the Magellan MIKE spectrograph. \toib is one of a few planets known to have a low-obliquity, high-eccentricity orbit, which may be the result of high-eccentricity coplanar migration. The planet has a periastron equilibrium temperature of $\sim 2100$\,K, similar to some highly irradiated hot Jupiters where atomic metal species have been detected in transmission spectroscopy, and varies by almost 1000 K during its orbit. Future observations of the atmosphere of TOI-2005b can inform us about its radiative timescales thanks to the rapid heating and cooling of the planet. 

\end{abstract}


\section{Introduction} \label{sec:Intro}

Unlike the case for hot Jupiters, the orbital eccentricities and inclinations of warm Jupiters, planets with $a/R_{\star} > 10$ and $R_{p} > 8\, R_{e}$, may preserve clues to the dynamical histories of close-in giant planets. This is because warm Jupiters experience relatively weak tidal interactions between the planet and its parent star, which would otherwise reduce the eccentricity and inclination. Thus, the orbital parameters of warm Jupiter systems are one avenue to decipher the mechanisms that drive the formation and orbital evolution of Jupiter systems. 

Only six transiting warm Jupiters with eccentricities larger than 0.5 have had their projected spin-orbit angles measured thus far (Figure\,\ref{fig:ecc_au}). Within this small sample, only two (TOI-1859b; \citealt{toi1859} and TOI-3362b; \citealt{toi3362,2023toi3362ecc}) orbit stars with $\teff > 6250$\,K. Early-type stars have little to no convective envelope and the obliquity of the system is therefore unlikely to be altered by tidal interactions \citep{winn2010}. Thus, when restricting attention to such hot stars, it may be easier to interpret the comparison between the spin-orbit angles of warm Jupiters and those of hot Jupiters, making obliquity measurements a particularly useful tracer for formation and evolution mechanisms. 

High-eccentricity tidal migration is thought to be the primary source of eccentric warm Jupiters which are thought to be planets that avoided tidal circularization via close periastron passages to form hot Jupiters \citep[see review by][]{DawsonJohnson2018}. Possible dynamical drivers include planet-planet scattering \citep[e.g.][]{1996Sci...274..954R, 2008Chatterjee,2008nagasawa}, secular interactions, such as von Zeipel-Lidov-Kozai oscillations \citep{1910zeipel, lidov1962,kozai1962} and secular chaos \citep{Wu2023}. These mechanisms all predict a wide distribution of orbital obliquities for the resulting planet population. 

In this paper, we report the planetary confirmation of \toib, a warm Jupiter on a 17.3-day eccentric orbit that is well aligned with the equatorial plane of a rapidly rotating F-star. In Section\,\ref{sec:Observations}, we describe the photometric data from \tess and multiple ground-based observatories, high-resolution speckle imaging, and CHIRON, FEROS, MINERVA-Australis, and Magellan-MIKE spectroscopic observations used to measure the stellar obliquity and to obtain an upper limit on the planetary mass. Section\,\ref{sec:global} describes the global modeling of the system to derive system parameters. We conclude with a discussion of the system and future atmospheric observational prospects in Section\,\ref{sec:discussion}.

\section{Observations} \label{sec:Observations}

\subsection{Photometric Observations} \label{subsec:photometry}
\subsubsection{TESS Photometry} \label{subsec:TESS}
\begin{figure*}
\begin{center}
\includegraphics[width=\textwidth]{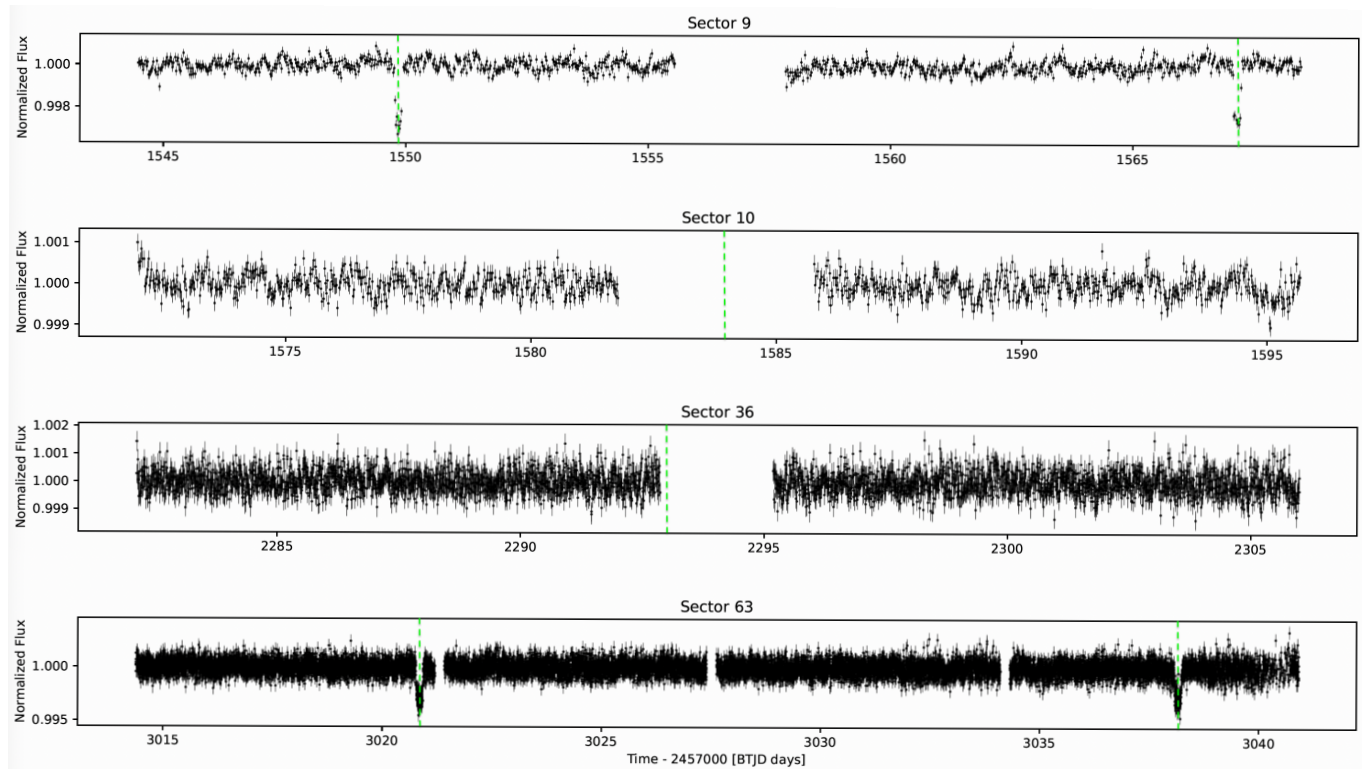}
\caption{Per-sector Normalized TESS PDCSAP light curves for TOI-2005. The target star was observed over four TESS sectors. The observing cadence was 1800 seconds in Sectors 9 and 10, 600 seconds in Sector 36, and 200 seconds in Sector 63. Unfortunately, during Sectors 10 and 36 the transit occurred in the gap in the lightcurve during data download. Predicted transits are marked with vertical dashed green lines.}
\label{fig:raw_tess}
\end{center}
\end{figure*}

Launched in April of 2018, the Transiting Exoplanet Survey Satellite (\textit{TESS}; \citealt{ricker2015}) is performing an all-sky survey in search of transiting exoplanets around nearby bright host stars. Each sector of observations is approximately 27 days in length and spans 24 $\times$ 96 degrees of the sky at a time. \toi (\tic) was observed by \textit{TESS} in Sectors 9 and 10 of Year\,1 of the primary mission with 1800-second cadence. It was then re-observed in the extended mission in Sector 36 during Year\,3 at 120-second and 600-second cadences and in Sector 63 during year 5 at a 20-second, 120-second, and 200-second cadences. The data were processed by the NASA Science Processing Operations Center pipeline (SPOC; \citealt{jenkins2016}) and the light curves were downloaded from the Mikulski Archive for Space Telescopes (MAST)\footnote{https://mast.stsci.edu/portal/Mashup/Clients/Mast/Portal.html} using the  \textit{Lightkurve} package \citep{lightkurve}. The Presearch Data Conditioning Simple Aperture Photometry (PDCSAP; \citep{2012PASP..124..985S,2014PASP..126..100S,2012PASP..124.1000S}) light curves were employed in our analysis and are plotted in Figure\,\ref{fig:raw_tess}. All TESS mission data used in this paper can be found in MAST: \dataset[https://doi:10.17909/fwdt-2x66]{https://doi:10.17909/fwdt-2x66}.

The SPOC \citep{jenkins2016} and QLP \citep{Huang2020,Kunimoto2022} pipelines identified a 17.3d candidate exoplanet orbiting \toi. The SPOC performed a transit search with an adaptive, noise-compensating matched filter \citep{Jenkins2002,Jenkins2010,Jenkins2020}, producing a Threshold Crossing Event (TCE) for which an initial limb-darkened transit model was fitted \citep{Li2019} and a suite of diagnostic tests were conducted to help assess the planetary nature of the signal \citep{Twicken2018}.  The transit signature was first detected in a search of Full Frame Image (FFI) data by the QLP pipeline at MIT. The QLP performed its transit search with the Box Least Squares Algorithm \citep{Kovacs2002}. The TESS Science Office (TSO) reviewed the vetting information and issued an alert on 17 June 2020 \citep{Guerrero2021}. The difference image centroid offsets localized the transit source for TOI 2005.01 within $2.4\pm2.5$ arcsec. 
The light curves were made available as part of SPOC processing of the TESS-SPOC science products \citep{caldwell2020}.

\subsubsection{Ground-based Photometry} \label{subsec:groundphot}

\begin{figure}
  \centering    \includegraphics[width=\linewidth]{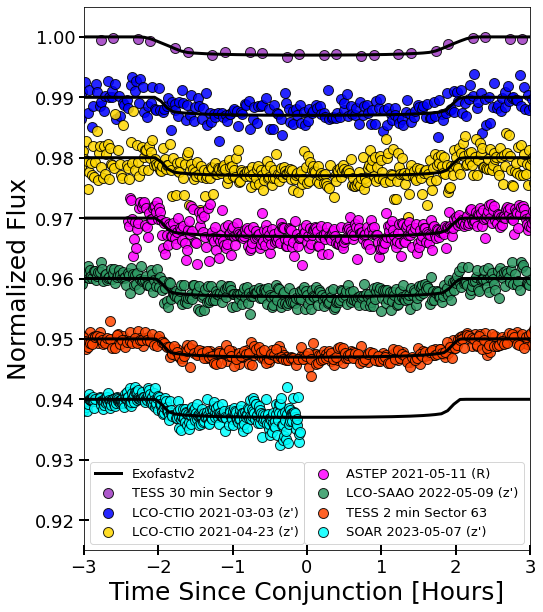} 
   \caption{Ground-based and TESS lightcurves normalized and offset relatively in flux and phased in time to the best-fit ephemeris. The best-fit model is shown in black.}\label{fig:allLCs}
\end{figure}

Follow-up ground-based photometric observations were performed by several observatories to aid in the confirmation of \toib as part of the TESS Follow-up Observing Program\footnote{https://tess.mit.edu/followup} Sub Group 1 \citep[TFOP;][]{TFOP}. On-target high resolution observations confirmed that the signal comes from the target and that there were no nearby eclipsing binaries that could potentially be contaminating the \tess photometry. The {\tt TESS Transit Finder}, a customized version of the {\tt Tapir} software package \citep{jensen2013}, was used to schedule ground-based transit observations. {\tt AstroImageJ} \citep[AIJ,][]{aij} was used for the data reduction and aperature phtometry unless otherwise noted. A short description of each facility is below and a summary of all the photometric observations can be found in Table\,\ref{tab:photometry}. The phase folded lightcurves are shown in Figure\,\ref{fig:allLCs}. All light curve data are available on the {\tt EXOFOP-TESS} website\footnote{\href{https://exofop.ipac.caltech.edu/tess/target.php?id=147660886}{https://exofop.ipac.caltech.edu/tess/target.php?id=147660886}} and are included in the global modeling described in Section\,\ref{sec:global}. Two datasets from the Perth Exoplanet Survery Satellite (PEST) Observatory were excluded from the global fits due to poor observing conditions. 

We observed two full transit windows of TOI-2005.01 in Pan-STARRS $z$-short band on UT 2021 March 03 and 2021 April 23 from the Las Cumbres Observatory Global Telescope \citep[LCOGT;][]{Brown:2013} 1\,m network node at Cerro Tololo Inter-American Observatory in Chile (CTIO). Another full transit window was observed in Pan-STARRS $z$-short band on UT 2022 May 09 from the LCOGT 1\,m network node at South Africa Astronomical Observatory near Sutherland, South Africa (SAAO). The 1\,m telescopes are equipped with $4096\times4096$ SINISTRO cameras with an image scale of $0\farcs389$ per pixel, resulting in a $26\arcmin\times26\arcmin$ field-of-view. The images were calibrated by the standard LCOGT {\tt BANZAI} pipeline \citep{McCully:2018}, and differential photometric data were extracted using {\tt AstroImageJ}. We used circular photometric aperture radii of $5\farcs8$, $6\farcs6$, and $5\farcs1$, respectively, that excluded all of the flux from the nearest known neighbor in the Gaia DR3 catalog (Gaia DR3 5388416255317688064) that is $10\farcs7$ east of TOI-2005. An $\sim$on-time $\sim$3 ppt event was detected in all three observations.

On UT 2021 May 11, a full transit of TOI-2005.01 was observed with ASTEP (Antarctic Search for Transiting ExoPlanets) \citep{Guillot+2015}, with a broad R filter between about 550 and 800\,nm \citep{Abe+2013} and 25 second exposures. Data processing was done as described in \cite{Mekarnia+2016}. The event was seen on time with a measured duration of 4:12 hours and a measured depth of $\sim2.6$ ppt for an uncontaminated 11.2" aperture radius, consistent with expectations from TESS. ASTEP also re-observed a full transit on UT 2022 April 22, this time using a new camera operating between 700 and 1000\,nm \citep{Schmider+2022} and 8 second exposures. The transit was again detected as predicted, with a similar depth and duration.

On UT 2023 May 07, we observed an ingress of TOI-2005.01 with the Goodman High Throughput Spectrograph (in imaging mode) on the SOuthern Astrophysical Research (SOAR) telescope \citep{Clemens:2004} at Cerro Tololo Interamerican Observatory in Chile. Because TOI-2005 is brighter in red wavelengths, we chose to utilize the Goodman Red camera for these observations, which is a $4096 \times 4112$ pixel charge-coupled device (CCD) detector with a spatial resolution of $0.15"/\text{pixel}$. Our observations were collected in the $z'$ filter with an exposure time of 11-seconds. The images were then calibrated and reduced using {\tt AstroImageJ}. No nearby known sources were included in the photometric aperture.

\begin{table*}
    \centering
    \caption{Ground-based Photometry}
    \label{tab:photometry}
    \begin{tabular}{lcccccl}
    \hline\hline
    \textbf{Observatory} & \textbf{Telescope size} &\textbf{Camera} & \textbf{Filter} & \textbf{Pixel Scale} & \textbf{UT Date} & \textbf{Detrend Parameters}\\
    & meters & & & arseconds & yyyy-mm-dd &\\
    \hline
LCO-CTIO & 1.0  & SINESTRO & $z^{'}$ & 0.39 & 2021-03-03 & focus position\\
LCO-CTIO & 1.0  & SINESTRO & $z^{'}$ & 0.39 & 2021-04-23 &  none \\
ASTEP & 0.4  & FLI Proline 16800E & $R_{c}$ & 0.93 & 2021-05-11 & airmass \\
LCO-SAAO & 1.0  & SINISTRO & $z^{'}$ & 0.39 & 2022-05-09 & sky/pixel, Y(fits)$_{T1}$ \\
SOAR & 4.1  & Goodman HTS & $z^{'}$ & 0.15 & 2023-05-07 & airmass, width$_{T1}$\\
\hline
    \end{tabular}
\end{table*}

\subsubsection{High-resolution Imaging} \label{subsec:imaging}

\begin{figure}
  \centering    \includegraphics[width=.9\linewidth,height=175pt]{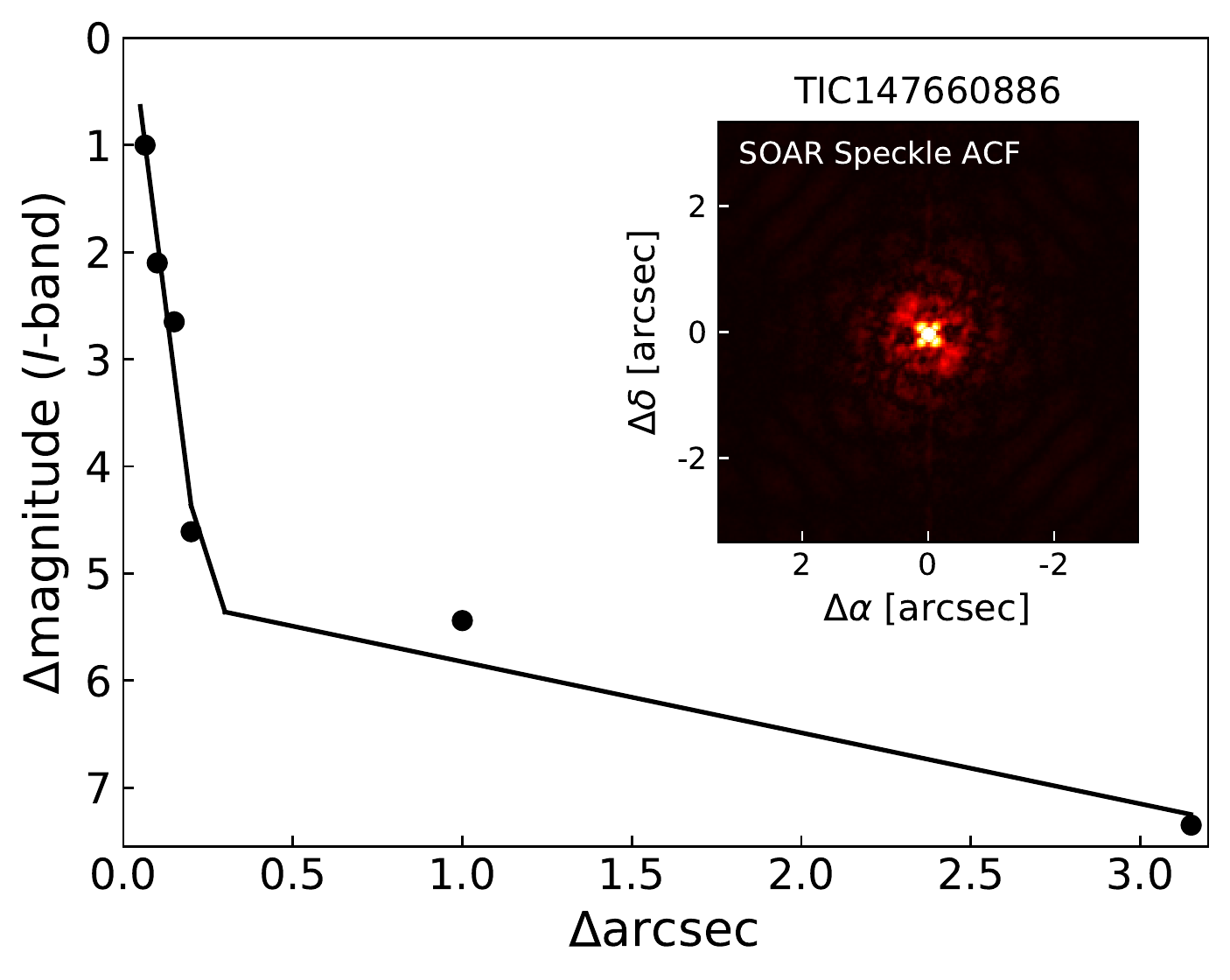} 
   \caption{ $5\sigma$ detection sensitivity and speckle auto-correlation functions of \toi from the 4.1-m SOAR. We find no companions to \toi within $3\arcsec$ in these high resolution imaging observations.}\label{fig:speckles}
\end{figure}

High-angular resolution imaging is needed to search for nearby sources that can contaminate the TESS photometry, resulting in an underestimated planetary radius, or be the source of astrophysical false positives, such as background eclipsing binaries. We searched for stellar companions to \toi with speckle imaging on the 4.1-m Southern Astrophysical Research (SOAR) telescope \citep{Tokovinin2018} on 31 October 2020 UT, observing in Cousins I-band, a similar visible bandpass as TESS. This observation was sensitive enough to detect a 5.6-magnitude fainter star at an angular distance of 1 arcsec from the target. More details of the observations within the SOAR TESS survey are available in \cite{ziegler2020}. The $5\sigma$ detection sensitivity and speckle auto-correlation functions from the observations are shown in Figure\,\ref{fig:speckles}. No nearby stars were detected within $3\arcsec$ of TOI-2005 in the SOAR observations.

\subsection{Spectroscopic Observations} \label{subsec:spec}
\begin{figure}
  \centering    \includegraphics[width=.9\linewidth,height=175pt]{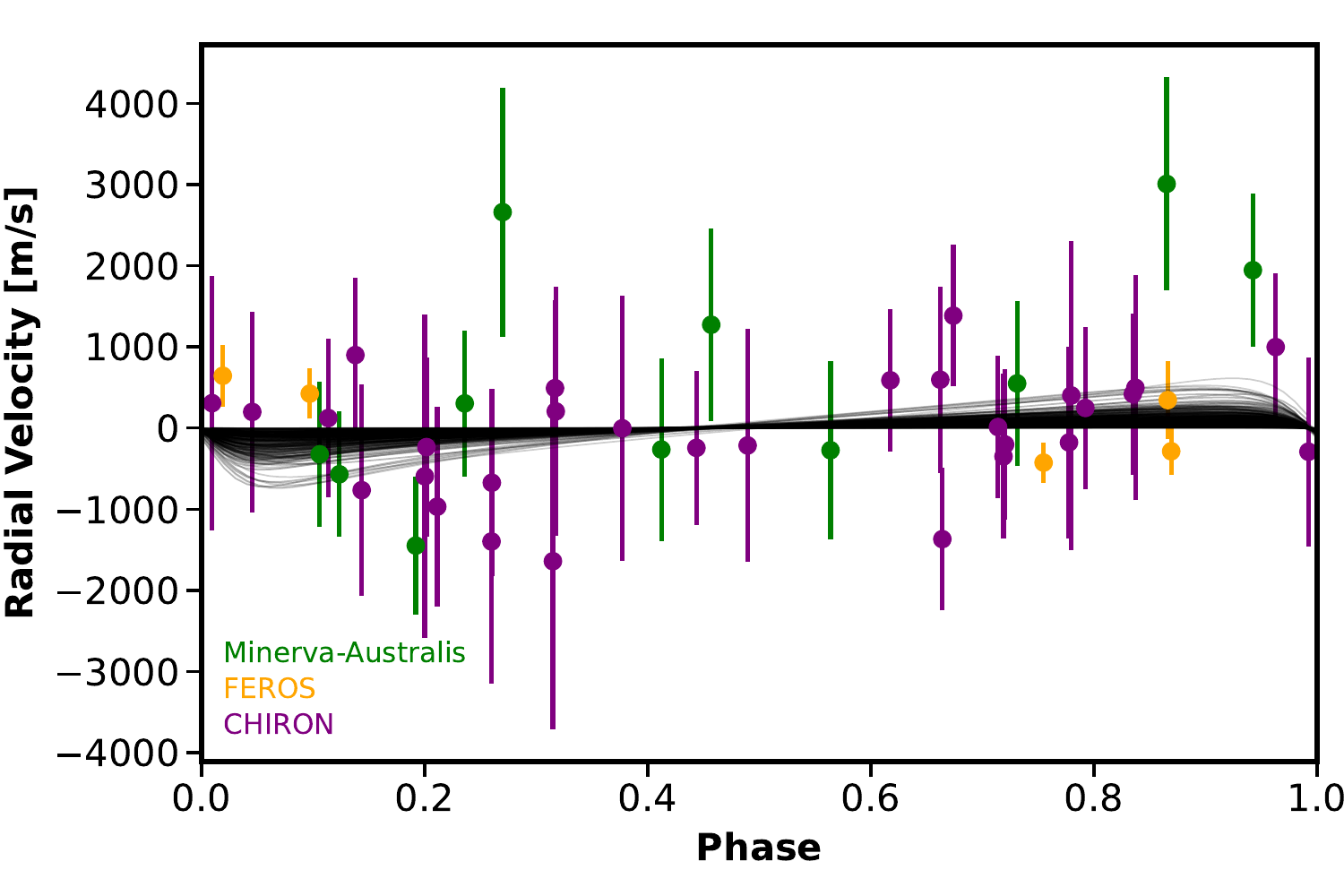} 
   \caption{Radial velocity observations phase folded to the ephemeris. Section \ref{subsec:spec} details the radial velocities observations. Over-plotted are 500 realizations of the converged models from the global modeling, illustrating the set of velocity orbits allowed by the transit and radial velocity observations, illustrating the constraint placed on eccentricity by the joint model. The $3\sigma$ radial velocity semi-amplitude of 1320 \ms\, provided a $3\sigma$ upper mass limit on the companion of $6.4 \mj$. }\label{fig:rvs}
\end{figure}

A total of 46 spectra were obtained from three separate observatories between UT 2020 March 20 and 2023 April 23. Details about the spectroscopic observations from each facility are described in more detail below. All radial velocities were derived via an analysis of the spectroscopic line profiles from least-squares deconvolutions. The observed spectra were deconvolved against a matching non-rotating synthetic template from the ATLAS9 model atmospheres \citep{Castelli:2004}. The resulting line profiles were modeled with a set of kernels representing contributions from rotation, macroturbulence, instrumental broadening, and radial velocity shift as per \citet{Gray:2005}. The velocity shifts applied to the kernels are the radial velocities of the star. These radial velocities are then combined order to order, as is standard for cross-correlation-function derived velocities, to determine the per-epoch velocity and uncertainty of the observation.
The radial velocities are listed in Table\,\ref{tab:rvs} and the phase folded velocities are shown in Figure\,\ref{fig:rvs}, along with 500 realizations of the converged models from Section~\ref{sec:global}. We also applied the Least Squares Deconvolution technique to the CHIRON spectra to determine the stellar rotational velocity. The analysis revealed an average \vsinistar of \vsiniLSD \kms. 


\begin{table}
    \centering
    \caption{Radial Velocities}
    \label{tab:rvs}
    \begin{tabular}{rrrc}
    \hline\hline
    \textbf{Time} & \textbf{Velocity} & \textbf{Uncertainty} & \textbf{Spectrograph}\\
    \textbf{[\bjdtdb]} & \textbf{[\kms]} & \textbf{[\kms]} & \\
    \hline
2458918.71188 & 7.89 & 2.07 & CHIRON \\
2458921.73230 & 9.32 & 1.44 & CHIRON  \\
2458925.61713 & 9.54 & 0.88 & CHIRON  \\
2459187.83937 & -6.35 &  0.48 & FEROS \\
2459191.83386 & -6.27 & 0.31 & FEROS \\ 
2459207.79145 & -6.05 & 0.38 & FEROS \\ 
2459209.84761 & 8.37 & 0.95 & CHIRON  \\   
2459242.85953 & 7.67 & 1.24 & CHIRON  \\  
2459245.72902 & 6.50 & 1.23 & CHIRON  \\  
2459249.75002 & 7.22 & 0.95 & CHIRON  \\  
2459252.75938 & 8.06 & 0.88 & CHIRON  \\  
2459253.73586 & 8.85 & 0.87 & CHIRON  \\  
2459255.78413 & 7.71 & 1.00 & CHIRON  \\ 
2459257.04496 & 12.33 & 1.31 & MINERVA-AUS \\
2459258.73708 & 8.46 & 0.91 & CHIRON  \\  
2459261.21232 & 9.00 & 0.89 & MINERVA-AUS \\
2459267.28583 & 10.59 & 1.19 & MINERVA-AUS \\
2459278.65059 & 7.59 & 0.98 & CHIRON  \\  
2459301.12396 & 9.06 & 1.12 & MINERVA-AUS \\
2460310.79654 & -7.12 & 0.25 & FEROS \\ 
2460312.77626 & -6.98 & 0.29  & FEROS \\
2459315.96764 & 11.98 &1.53 & MINERVA-AUS \\
2459323.94920 & 9.87 & 1.01 & MINERVA-AUS \\
2459328.46905 & 7.18 & 1.17 & CHIRON  \\ 
2459328.76576 & 7.77 & 1.57 & CHIRON  \\  
2459331.92638 & 7.87 & 0.85 & MINERVA-AUS \\
2459344.91279 & 11.27 & 0.94 & MINERVA-AUS \\
2459348.04569 & 8.75 & 0.77 & MINERVA-AUS \\
2459349.99132 & 9.62 & 0.89 & MINERVA-AUS \\
2459372.97197 & 9.05 & 1.10 & MINERVA-AUS \\
2459988.71106 & 8.77 & 1.30 & CHIRON  \\
2459989.71543 & 9.30 & 1.10 & CHIRON  \\
2459990.72537 & 8.14 & 1.75 & CHIRON \\
2459991.72212 & 9.74 & 1.53 & CHIRON \\
2459992.75463 & 9.53 & 1.63 & CHIRON  \\
2459997.68743 & 10.13 & 1.15 & CHIRON  \\
2459998.69250 & 9.33 & 0.92 & CHIRON  \\
2459999.71994 & 9.93 & 1.91 & CHIRON  \\
2460000.71447 & 10.03 & 1.39 & CHIRON  \\
2460041.60773 & 8.94 & 1.99 & CHIRON  \\
2460042.64769 & 8.86 & 1.15 & CHIRON  \\
2460043.62817 & 10.02 & 1.09 & CHIRON  \\
2460049.63675 & 8.16 & 0.88 & CHIRON  \\
2460050.58596 & 9.18 & 1.02 & CHIRON  \\
2460051.60215 & 9.35 & 1.18 & CHIRON  \\
2460052.59233 & 9.95 & 0.99 & CHIRON  \\
\hline
    \end{tabular}
\end{table}

\subsubsection{CHIRON} \label{subsec:Chiron}
CHIRON is a fiber-fed echelle spectrograph on the 1.5m SMARTS telescope at Cerro Tololo Inter-American Observatory, Chile \citep{2013CHIRON}. CHIRON observes over the wavelength range $4100-8200\, \AA\,$ and has a spectral resolving power of $R\sim 80,000$. Spectra were extracted as per the official CHIRON pipeline \citep{2021CHIRONreduction}. Thirty observations were obtained between March 2020 and April 2023. 

\subsubsection{FEROS} \label{subsec:feros}
The Fiber-fed Extended Range Optical Spectrograph \citep[FEROS][]{feros} is a high resolution, temperature stabilized echelle spectrograph installed at the MPG-2.2m telescope, in the ESO La Silla Observatory, in Chile. FEROS has a resolving power of 48,000 and uses a second fiber to trace instrumental wavelength shifts. We obtained 5 FEROS spectra between between December 2020 and January 2024 in the context of the Warm gIaNts with tEes collaboration \citep[WINE][]{brahm:2019, jordan:2020,hobson:2023,jones:2024}. The adopted exposure time was 300-second and the typical signal-to-noise ratio was of 120. All reduction and processing steps were performed using the \texttt{ceres} pipeline \citep{ceres}.

\subsubsection{MINERVA-Australis} \label{subsec:Minerva}
MINERVA-Australis is an array of telescopes located at Mt Kent Observatory, Australia. It includes four identical 0.7\,m telescopes feeding into one Kiwispec echelle spectrograph \citep{2019PASP..131k5003A}. Eleven observations for \toi{} were performed between February and July 2021. Radial velocities were derived using the least-squares deconvolution fit from each telescope individually. The velocities were weighted and combined after removing a fitted-for offset between each telescope. The combined velocities are presented in Table\,\ref{tab:rvs}.

\subsubsection{Transit Spectroscopic Observation} \label{subsec:DT}
Spectroscopic observations were obtained during transit to measure the projected spin-orbit angle of the system. This technique is based on a phenomenon called the Rossiter-McLaughlin or RM effect \citep{rossiter,mclaughlin}, in which the partial blockage of the rotating stellar surface by the transiting planet produces a distortion in the star's spectral absorption lines. For example, when the planet is projected in front of the receding half of the star, some of the redshifted components of the line profiles are missing, and the overall line profile appears to be blueshifted.

The Magellan Inamori Kyocera Echelle (MIKE) spectrograph \citep{MIKE} on the Magellan Telescope at Las Campanas Observatory (LCO) in Chile was used to obtain simultaneous data with the blue (320-480 nm) and red (440-1000 nm) channels of the spectrograph on the night of 2023 Feb 27. Using a 0.35 arcsecond slit, MIKE-blue has a resolving power of $R=83{,}000$ and MIKE-red has a resolving power of $R=65{,}000$. Observations were acquired in sets of six 600-second observations alternating between the blue and red channels. Two Thorium-Argon calibration spectra were obtained between every set of observation. Calibrations and spectral extraction via the \textsc{carpy} package \citep{2000ApJ...531..137K,2003PASP..115..688K}. The seeing was variable but mostly better than 1 arcsecond during the observations. 

\begin{figure*}
     \centering
\begin{tabular}{cc}
\textbf{MIKE Blue} & \textbf{MIKE Red}  \\
  \includegraphics[width=0.8\columnwidth]{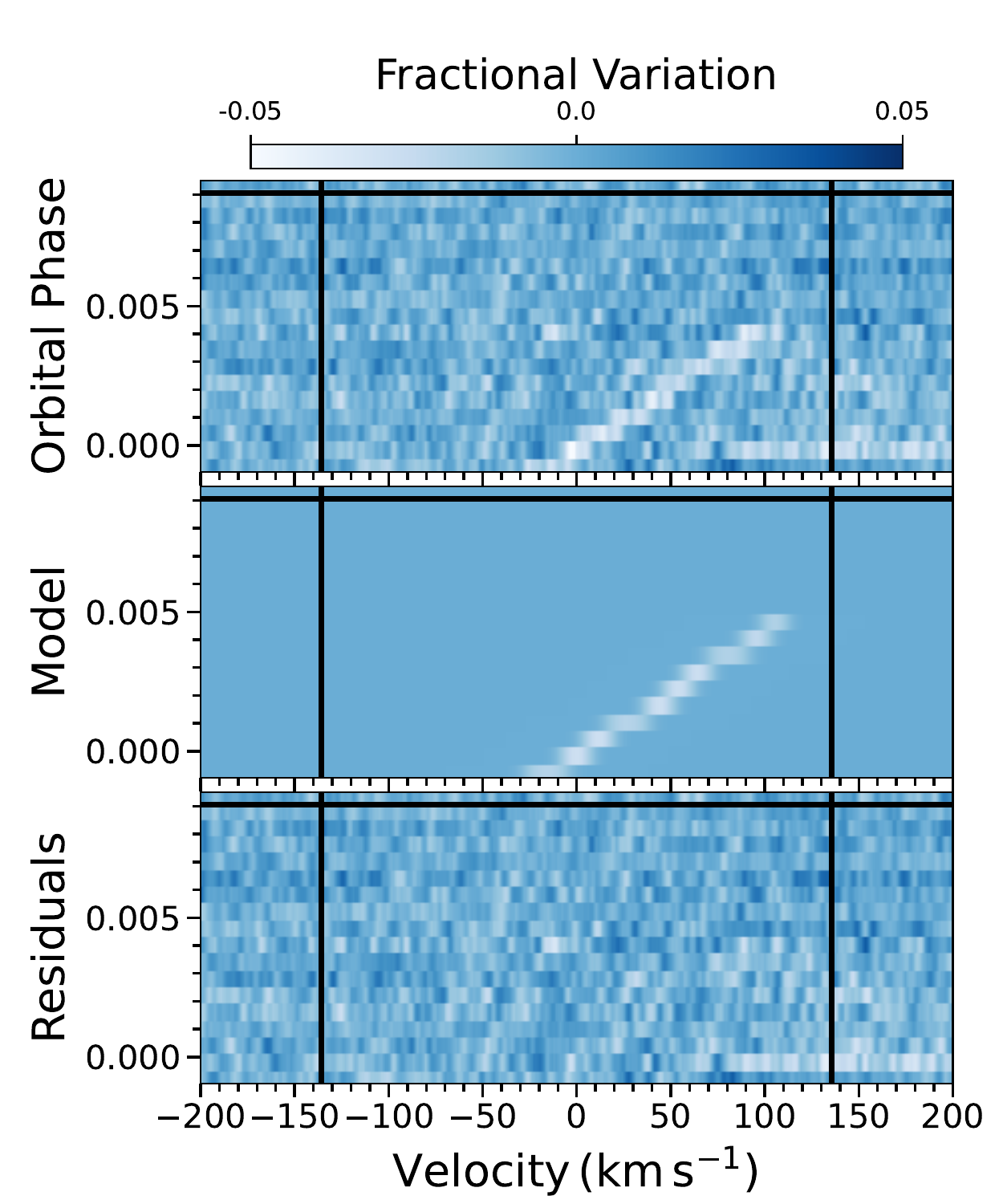} &
  \includegraphics[width=0.8\columnwidth]{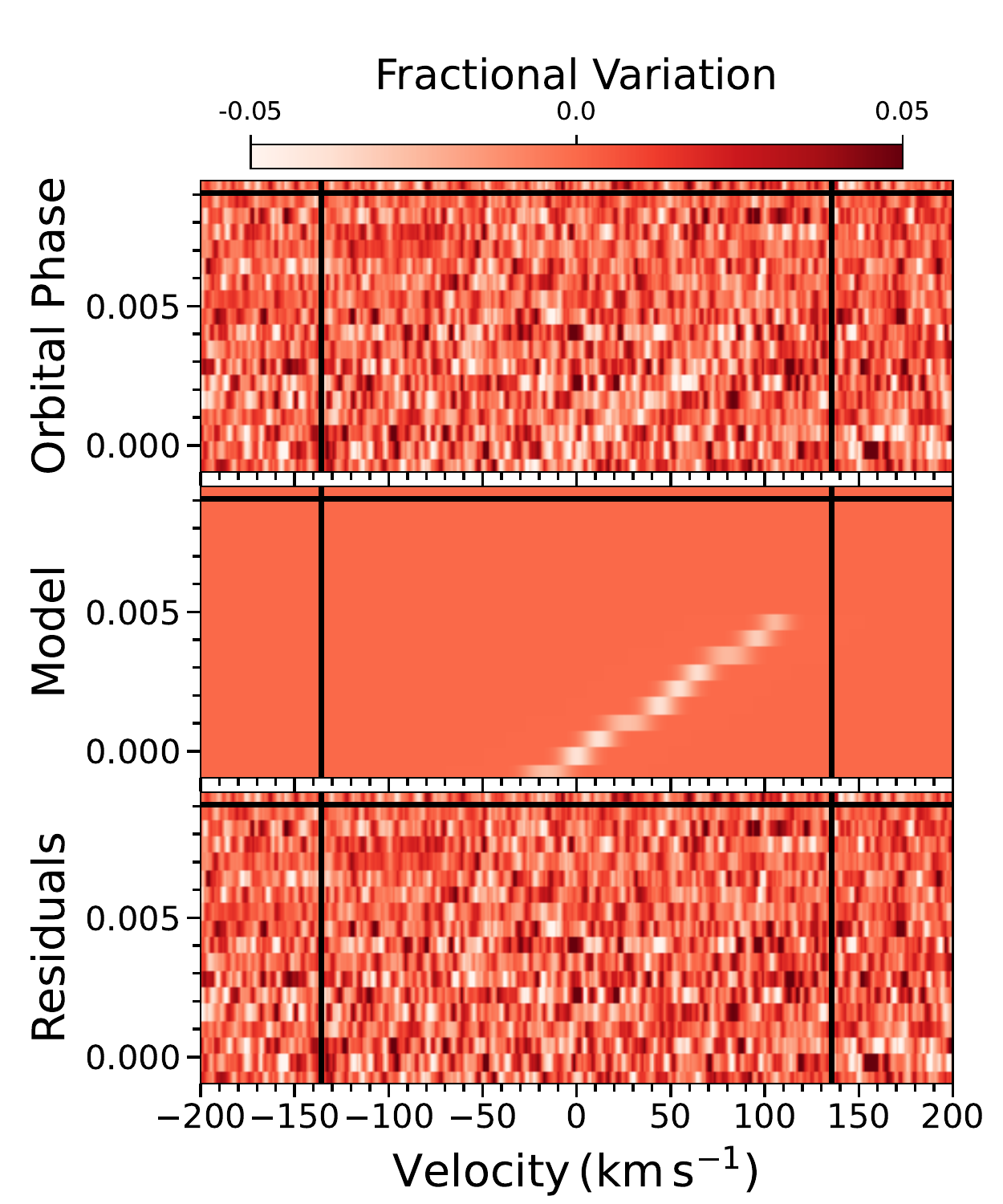} \\
\end{tabular}
  \caption{MIKE Doppler spectroscopic results from February 27, 2023. The color scale represents the fractional variation in the spectral line profile. Each plot shows the planetary signal after the average rotational profile is subtracted (top), the best-fit model (middle), and the residuals after subtracting the planetary signal (bottom).}\label{fig:DT}
\end{figure*}

The Doppler shadow is extracted by deriving the line broadening profile using a least squares deconvolution between the observed spectra and a synthetic stellar template
spectrum. The template spectrum was based on ATLAS9 \citep{Castelli:2004} and was generated based on the estimated atmospheric parameters of the host star, with no rotational broadening incorporated. The portion of the star blocked by the transiting planet is modeled by the difference between the derived line profile and the median combined line profile, as described in \citet{2019AJ....157...31Z}. The analysis incorporates the effects of local limb darkening, macroturbulent broadening, and rotation. The line profile residuals for each transit observation, and for the combined observations, are shown in Figure~\ref{fig:DT}. The light trail from bottom left to top right represents the shadow of the planet during the transit in the line profile residuals. 

The Doppler analysis was checked using two independent pipelines. In the first approach, we performed a joint fit of the TESS photometry and the Doppler shadow. Each TESS transit was modeled with a standard transit model plus a Gaussian Process model, described by the Matérn 3/2 kernel, for the treatment of correlated noise. For the Doppler shadow modeling, we first calculated the planet's position on the stellar disk, and the corresponding stellar rotation velocity being blocked from view. We then convolved this ``subplanet spectrum'' with a Gaussian velocity profile. The width of the Gaussian profile was an adjustable parameter that depends on the resolution of the spectrograph and the macroturbulence of the star. Lastly, we calculated the likelihood by comparing the normalized planetary velocity profile at each time with the observed Doppler shadow. A detailed description of the model is provided in Section 4.2 of \cite{Dong22}.
The model is built using the $\mathtt{exoplanet}$ and $\mathtt{PyMC}$ packages. After optimizing to determine the initial point for the Markov Chain Monte Carlo (MCMC), we ran 4 chains, each with 5000 tuning steps and 3000 draws. A target acceptance rate of 0.95 was used to avoid divergences.

We also performed a full global analysis using all available photometric and spectroscopic data as described in Section\,\ref{sec:global} and adopted the results as our final parameters as shown in Table\,\ref{tab:bestfit}.

\subsection{SED analysis} \label{sec:sed}
\begin{figure}
    \centering    \includegraphics[width=0.46\textwidth]{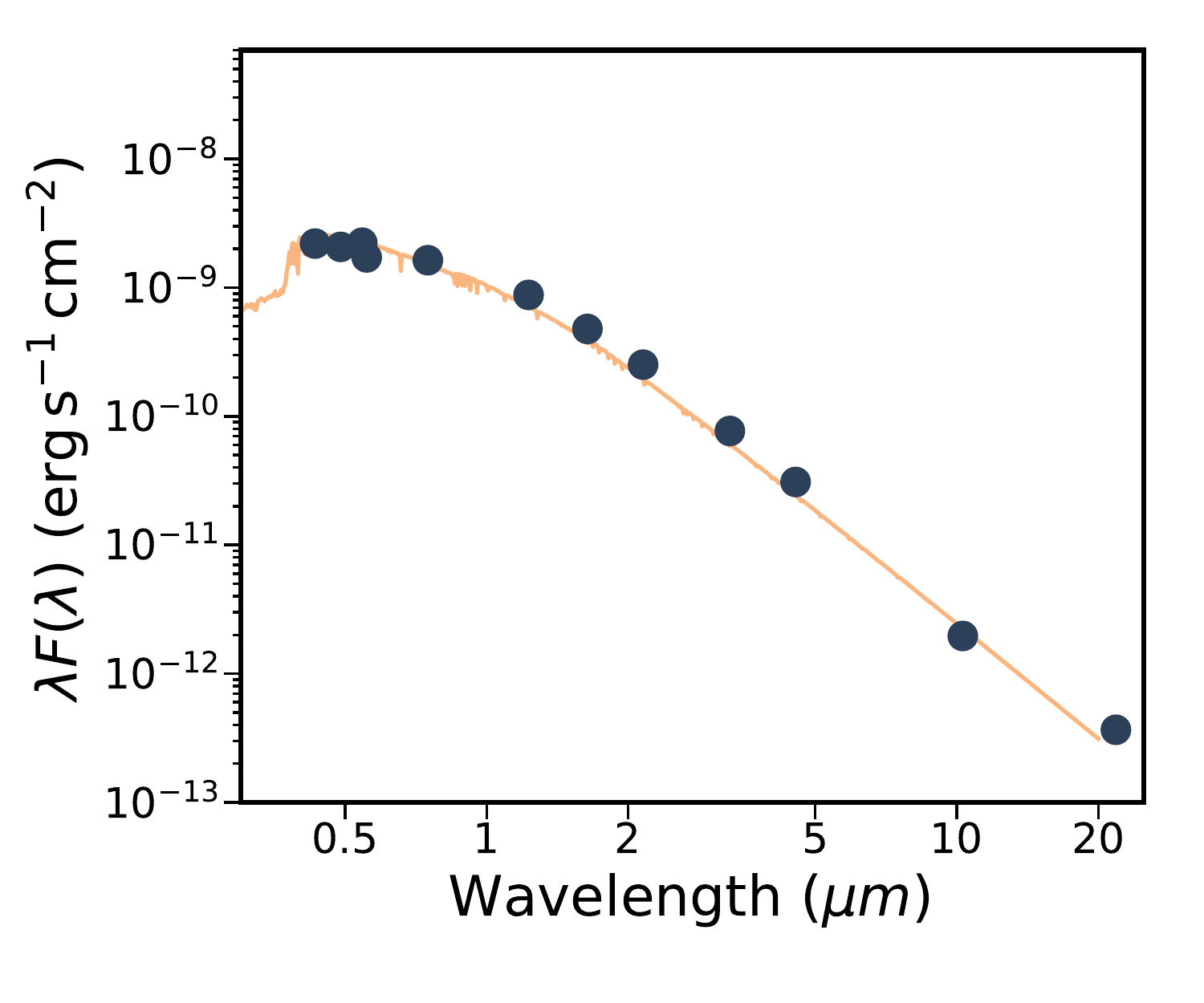}
    \caption{Spectral energy distribution of the target star \toi. Magnitudes from \emph{Gaia} $G$, $B_p$, $R_p$, \emph{Tycho} $B$, $V$, 2MASS $J$, $H$, $Ks$, and WISE $W1$, $W2$, $W3$, and $W4$ are included in the global modeling of the system and are shown in dark blue. }\label{fig:sed}
\end{figure}

We used all available broadband photometry, including \emph{Hipparcos} $B$ and $V$ bands \citep{1997AA...323L..49P}, \emph{Gaia} DR3 $G$, $Bp$, $Rp$ \citep{2022arXiv220800211G}, 2MASS $J$, $H$, $K$ \citep{2006AJ....131.1163S}, and WISE $W_1$, $W_2$, $W_3$, $W_4$ bands \citep{WISE}, as well as \emph{Gaia} DR3 parallaxes to construct the spectral energy distribution of \toi. The spectral energy distribution was modeled simultaneously with the photometric and spectroscopic observations of the system which is discussed in the next Section. The best-fit model is shown in Figure\,\ref{fig:sed} and the catalog parameters are reported in Table\,\ref{tab:star}. 

\begin{table}
    \caption{Literature values for \toi.
    \tablewidth{\columnwidth}
    \label{tab:star}}
    \centering
    \begin{tabular}{lll}
    \hline\hline
\textbf{Stellar Parameters} & \textbf{Value} & \textbf{Source}\\
\hline
\multicolumn{3}{l}{\textbf{Catalog Information}}\\
TIC ID & 147660886 &   TOI Catalog \\
TOI ID & 2005 &  TOI Catalog \\
\textit{Gaia} DR3 ID & 5388416255319817856 & GAIA DR3\\
\textit{2MASS} ID & J10560763-4322362 & 2MASS\\
TYC ID & 7736-00803-1 & Tycho\\
\multicolumn{3}{l}{\textbf{Coordinates and Proper Motion}} \\
Right Ascension (h:m:s)  & 10:56:07.60 (J2000) &  GAIA DR3 \\
Declination (d:m:s)    & -43:22:36.1 (J2000) &  GAIA DR3 \\
Parallax (mas)   & \parallax &  GAIA DR3 \\
$\mu$\textsubscript{R.A} (mas yr\textsuperscript{-1}) & \pmRA & GAIA DR3 \\
$\mu$\textsubscript{Dec.} (mas yr\textsuperscript{-1}) & \pmDEC &  GAIA DR3 \\
\multicolumn{3}{l}{\textbf{Magnitudes}} \\
$TESS$ (mag)         & $9.492 \pm 0.006$ &  TOI Catalog \\
$G$ (mag)         & 9.76491~$\pm$~0.00013 & GAIA DR3 \\
$B_{p}$ (mag)         & 9.95198~$\pm$~0.00018 & GAIA DR3 \\
$R_{p}$ (mag)         & 9.43687~$\pm$~0.00017 & GAIA DR3 \\
$B$ (mag)      & 10.222~$\pm$~0.049 & Tycho \\
$V$ (mag)         & 9.867~$\pm$~0.003 & Tycho \\
$J$ (mag)         & 9.099~$\pm$~0.034 & 2MASS \\
$H$ (mag)         & 8.977~$\pm$~0.023 & 2MASS \\
$K$ (mag)        & 8.905~$\pm$~0.019 & 2MASS\\
$WISE_{3.4\mu}$ (mag)     & 8.863~$\pm$~0.022 & WISE \\
$WISE_{4.6\mu}$ (mag)     & 8.882~$\pm$~0.020 & WISE \\
$WISE_{12\mu}$ (mag)     & 8.846~$\pm$~0.026 & WISE \\
$WISE_{22\mu}$ (mag)     & 8.663~$\pm$~0.348 & WISE \\
\hline
\end{tabular}
\begin{flushleft}
\footnotesize{\textbf{Note:} TESS TOI Primary Mission Catalog; \citep{Guerrero2021}, Tycho; \citep{tycho}; GAIA DR3; \citep{Gaiadr3}, 2MASS; \citep{2MASS}, WISE; \citep{allwise}}
\end{flushleft}
\end{table}

\section{Global Modeling} \label{sec:global}

We performed a joint analysis of all available photometric, spectroscopic, and catalog observations to determine the stellar and planetary parameters of the system. This included photometric transit observations from TESS and follow-up ground-based observatories (Section\,\ref{subsec:TESS} and \ref{subsec:groundphot}), a spectroscopic transit observation from MIKE (Section\,\ref{subsec:DT}), and radial velocities from CHIRON, FEROS, and MINERVA-Australis (Section\,\ref{subsec:spec}). Free parameters largely describing the transit include the orbital period $P$, reference time of transit center $T_0$, planet-to-star radius ratio $R_p/R_\star$, line-of-sight inclination $i$, and orbital eccentricity parameters $\sqrt{e}\cos \omega$ and $\sqrt{e} \sin \omega$ where $e$ is the orbital eccentricity and $\omega$ the argument of periastron. In addition, the radial velocity orbit and spectroscopic transits are described by the mass of the planetary companion $M_p$ and its projected orbital obliquity $\lambda$. Simultaneous with the transit models, we also interpolated the stellar isochrones as per Section\,\ref{sec:sed}. At each step, we modeled the spectral energy distribution to constrain the stellar parameters. Free parameters describing the stellar isochrone and spectral energy distribution modeling include stellar mass $M_\star$, age, metallicity [m/H], and parallax. Parallax is tightly constrained by a Gaussian prior about its \emph{Gaia} DR3 value and associated uncertainties. We also include free parameters for the rotational and macroturbulent velocities to enable modeling of the spectroscopic transits. These values are also tightly constrained by their spectroscopic measurements and associated uncertainties. 

The photometric transits are modeled as per \citet{mandelAgol2002} via the \textsc{batman} package \citep{2015PASP..127.1161K}. Limb darkening parameters are interpolated and fixed to their values as per \citet{claret2011}, \citet{2017A&A...600A..30C} and \citet{eastman2013}. The spectroscopic transit is modeled as per Section~\ref{subsec:DT}, via a disk integration of the portion of the stellar surface occulted by the planet, incorporating the effect of local macroturbulence and rotational broadening. The best fit parameters are presented in Table\,\ref{tab:star} and Table\,\ref{tab:bestfit}.

\begin{deluxetable*}{llrr}
\tablecaption{Best-fit Stellar and Planetary Properties for \toi   \label{tab:bestfit}}
\tablecolumns{4}
\tablehead{
\colhead{Parameters} &
\colhead{Description (Units)} & \colhead{Prior Values} &
\colhead{Best Fit}
}
\startdata
\textbf{Stellar Parameters:} \\
$M_{\star}$ & Stellar Mass ($M_\odot$) & $\mathcal{U}(1,3)$ & \mstar \\
$R_{\star}$ & Stellar Radius ($R_{\odot}$) & Inferred & \rstar  \\
$L_{\star}$ & Stellar Luminosity ($L_{\odot}$) & Inferred & \lstar \\
$T_{\rm eff}$ & Effective Temperature (K)  & Inferred & \tefffit \\
$\log g$ & Surface Gravity (cgs)   & Inferred & \loggfit \\
$\mathrm{[m/H]}$ & Metallicity (dex) & Fixed & 0 \\
$v \sin i$ & Projected Rotational Velocity (\kms)  & $\mathcal{G}(110,1.0)$ & \vsiniLSD   \\
$v_\mathrm{macro}$ & Macroturbulent velocity (\kms) & $\mathcal{U}(0,20)$ & $3.7 \pm 1.6$ \\
Parallax & Parallax (mas) & $\mathcal{G}(11.409,0.026)$ & $11.409 \pm\, 0.025$ \\
Age & Age (Gyr) & $\mathcal{U}(0.1,2.0)$ & \age \\
Distance & Distance (pc) & Inferred & \dist\\
$E(B-V)$ & Reddening (mag) & Inferred & $0.012^{+0.013}_{-0.009}$\\
\hline
\textbf{Planetary Parameters:} \\
$P$ & Orbital Period (days) & $\mathcal{U}(17.2,17.4)$ & \per \\
$T_{o}$ & Epoch (BJD) & $\mathcal{U}(2458549.8,2458549.9)$ & $2458549.8376_{-0.0013}^{+0.0011}$ \\ 
$M_{p}$ & Planet Mass (\mj) & $\mathcal{U}(0,100)$ & $< 6.4 \, (3\sigma)$ \\
$R_{p}$ & Planet Radius (\rj) & Inferred & \plrad \\
$R_{p}/R_{\star}$ & Radius of planet to star ratio  & $\mathcal{U}(0.03,0.10)$ & $0.05107^{+0.00037}_{-0.00045}$ \\
$a/R_{\star}$ & Semi-major axis to star radius ratio  & Inferred & $15.9^{+1.97}_{-1.75}$ \\
$a$ & Semi-major axis (AU) & Inferred & \semimaj \\
$e$ & eccentricity  & Inferred & $0.597_{-0.065}^{+0.097}$  \\
$\sqrt{e} \cos \omega$ & eccentricity parameter & $\mathcal{U}(-1,1)$\tablenotemark{a} & $-0.254_{-0.078}^{+0.065}$ \\
$\sqrt{e} \sin \omega$ & eccentricity parameter & $\mathcal{U}(-1,1)$\tablenotemark{a} & $0.729_{-0.040}^{+0.024}$ \\
$\gamma$ & Radial velocity offset (\kms) & $\mathcal{U}(24,25)$ & $24.424_{-0.072}^{+0.077}$\\
$T_{14}$ & Transit duration (hours) & Inferred & $2.249 \pm 0.063$ \\
$K$ & RV semi-amplitude (m/s)  & Inferred & $< 1320\, (3\sigma)$\\
$i$ & Transit inclination (deg) & $\mathcal{U}(80,90.5)$\tablenotemark{a} & $87.1^{+1.1}_{-0.9}$ \\
$b$ & Impact parameter & Inferred & $0.52_{-0.35}^{+0.33}$ \\
$\lambda$ & Projected Spin-Orbit Angle (deg) & $\mathcal{U}(0,180)$ & \lam \\
\enddata
\tablenotetext{a}{Parameters that result in $e>1$ or non-transiting solutions are excluded within the MCMC.}
\tablenotemark{}{Note: Inferred parameters are not directly modeled but are derived from other parameters.}
\end{deluxetable*}

\section{Discussion} \label{sec:discussion}

\toib is a warm Jupiter with an orbital period of 17.3 days. We were unable to detect an orbital signal due to the rapid rotation of the host star, but we have placed a $3\sigma$ upper limit on the planet mass of $6.4\,\mj$ and determined that the object is planetary in nature. The planet orbits an early F-type star (\teff = \tefffit\,K) with a mass of $1.59\,\msun$ and a radius of \rstar\, \rsun. The planet orbits the host star in an eccentric orbit determined to be \ecc\, with a measured sky-projected spin-orbit angle of $\lambda =\,$\lam\ deg.

\begin{figure*}
\centering
    \includegraphics[width=0.8\textwidth]{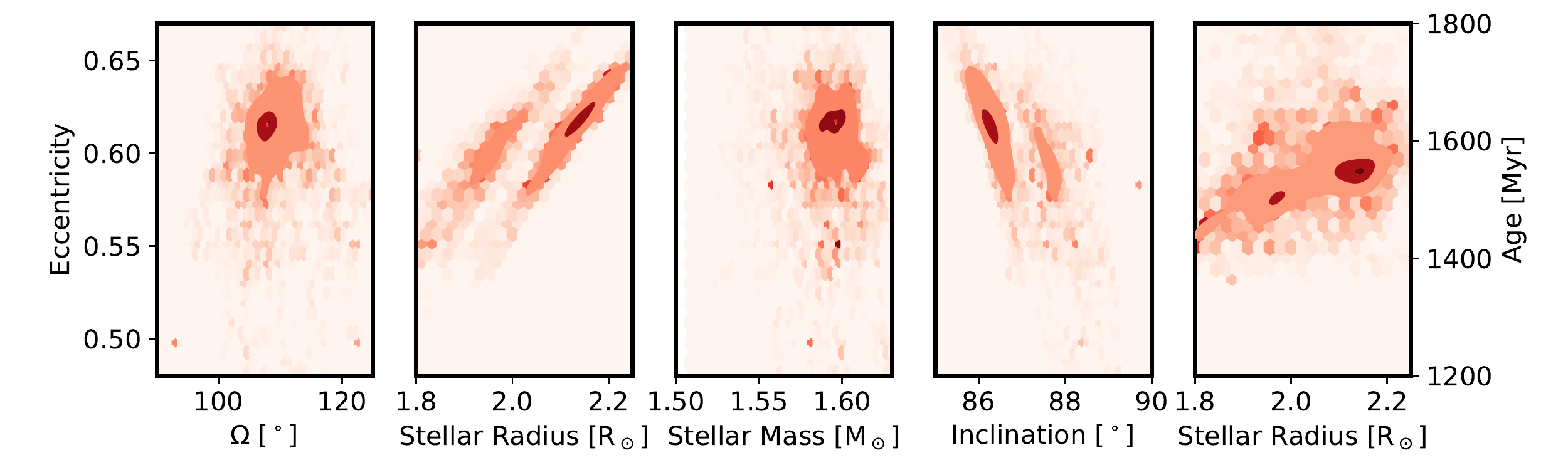}
    \caption{Correlations between eccentricity and key parameters that affect its posterior distribution, including the argument of periastron $\Omega$, stellar mass and radius, and transit inclination. The $1\sigma$ and $2\sigma$ regions are marked by the red 2D histograms. We note a bimodality in the stellar radius as the posterior overlaps between the main sequence and subgiant branches. }\label{fig:ecc_corner}
\end{figure*}

\begin{figure*}
\begin{center}
\includegraphics[width=\textwidth]{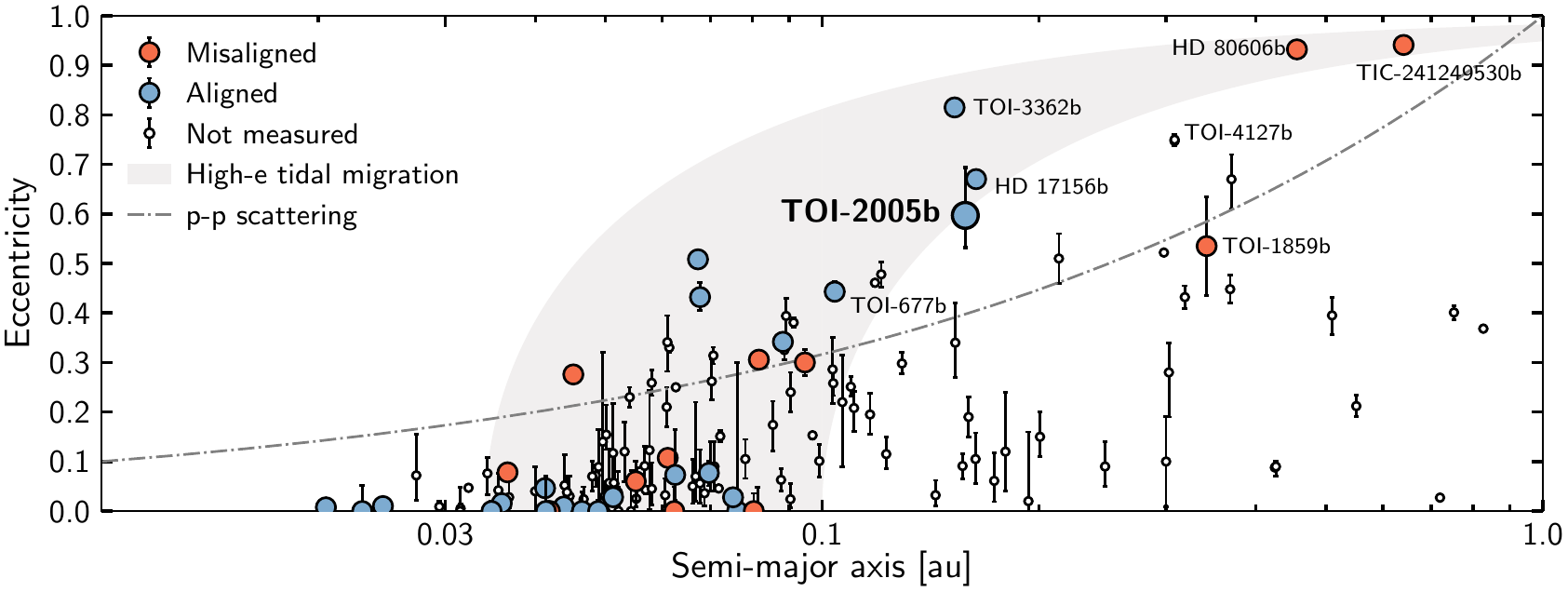}
\caption{Eccentricity vs. semi-major axis for all confirmed transiting planets between 6--20 Earth radii. Colored circles are planets with measured spin-orbit angles. Blue circles are defined as `aligned' with spin-orbit angles of less than 30 degrees and red circles are defined as `misaligned' with spin-orbit angles greater than 30 degrees. The gray region illustrates planets that are likely undergoing high-eccentricity tidal migration. The upper and lower limits of the track are set by the Roche limit and the tidal circularization timescale \citep{toi3362}. \toib sits on the lower edge of the migration track. The dotted-dashed line presents the theoretical upper limit of eccentricities as a result of planet-planet scattering, assuming a planet with a mass of 0.5 $M_{\rm Jup}$ and a radius of 2 $R_{\rm Jup}$, for illustrative purpose \citep{Petrovich:2014}.
Data were compiled from the NASA Exoplanet Archive as of October 30, 2024. Stellar obliquity measurements were obtained from \cite{Knudstrup24} Table B2 with recent updates.} 
\label{fig:ecc_au}
\end{center}
\end{figure*}

\subsection{High eccentricity coplanar migration}

The highly eccentric nature of \toib is determined via a joint modeling of the photometric transit and the radial velocity orbit. The stellar parameters are well constrained by the spectral energy distribution and \emph{Gaia} parallax, providing a constrained stellar mass and radius. The short transit duration of \toib, compared to that expected for a circular orbit, is indicative that the transit is occurring at periastron passage. Higher eccentricity solutions are constrained by the lack of a radial velocity detection at periastron. The posterior distribution for eccentricity against relevant parameters that influence its value and precision are shown in Figure~\ref{fig:ecc_corner} for reference.

\toib might be undergoing high-eccentricity tidal migration as inferred from the semi-major axis and eccentricity, and shown in Figure \ref{fig:ecc_au}. \toib currently resides at a semi-major axis of \semimaj\,au but as the planet passes through periastron on its eccentric orbit, energy dissipation interior to the planet will lead to tidal circularization. The planet has an eccentricity of \ecc\, and a semi-major axis ($a_{\textnormal{p}}$) of \semimaj\,au. If its orbit can be circularized within the life of the planetary system, it will end up with an orbital distance of 0.101\,au, as defined by $a_{final} = a(1-e^{2})$ following the evolution of constant angular momentum tracks (see \cite{DawsonJohnson2018} for a review). The timescale for tidal circularization varies and relies heavily on the quality dissipation parameter, $Q'_{p}$. If we calculate the tidal circularization for the system (Equation \ref{eq:tcirc}) adopted from \cite{AdamsLaughlin2006} while assuming a $Q'_{p} = 10^6$, we calculate that the circularization timescale of the planet is $\sim 10^{11}$ years and much longer than the age of the system. 
\begin{equation}
\label{eq:tcirc}
    \tau_{circ}\sim\frac{4Q'_{p}}{63}(\frac{a}{GM_{*}})^{1/2}\frac{M_{p}}{M_{*}}(\frac{a}{R_{p}})^{5}(1-e^{2})^{13/2}[F(e^{2})]^{-1}
\end{equation}
We note that the circularization timescale would change depending on what $Q'_{p}$ value is adopted and that $Q'_{p}$ does not stay constant during the planet's evolution. 

Most high eccentricity migration mechanisms result in a wide range of final orbital obliquities \citep[see summaries in][]{albrecht2022}. One mechanism for hot Jupiters existing with high eccentricity and low obliquity is explained through Coplanar High-Eccentricity Migration \citep{2015Petrovich}. In this scenario, a coplanar inner Jovian planet is formed via secular interactions between it and another giant planet much farther out at low mutual inclinations. The farther out companion would be difficult to detect given the long period, the small RV signal, and the host star's rotational velocity. We do not detect any additional bodies in our current dataset. Coplanar high-eccentricity migration is applicable to some outcomes of prior planet-planet scattering events that result in high eccentricity orbits for both the inner and outer planet at low mutual inclinations \citep{2013AJ....146...63T}. Coplanar high-eccentricity migration then acts to migrate the inner Jovian planet inward via high eccentricity and tidal circularization, while keeping the mutual inclination of the two planets low. In many cases, this also results in the formation of a coplanar, low obliquity hot Jupiter. Another method for generating low or moderately eccentric Jupiters is through disk-driven migration. Typically disk-driven migration is thought to be a smooth migration mechanism but if the planet migrates into a low-density disk cavity, the eccentricity can increase due resonances and will not be opposed given the lack of gas in the cavity \citep{2021MNRAS.500.1621D, 2023ApJ...956...17L}. 

The discovery of low-obliquity, eccentric warm Jupiters can help trace the prevalence of this migration pathway. Previous examples include HD\,17156 b \citep{2007ApJ...669.1336F,2009A&A...503..601B}, TOI-3362 b \citep{toi3362,Espinoza-Retamal:2023},  HD 118203 b \citep{2020AJ....159..243P,Zhang:2024}, and TOI-677 b \citep{2020AJ....159..145J,Sedaghati:2023}. In fact, of the eight planets with obliquities measured that are clearly within the high-eccentricity evolution track, and exhibit modest eccentricities larger than that predicted by planet-planet scattering (see the dotted-dashed line in Figure\,\ref{fig:ecc_au}), six have low projected obliquities. Assuming these projected obliquities are representative of their true 3D obliquities, their predominance is a clue that co-planar migration plays an important role in the formation of hot Jupiters.  

\subsection{Atmospheric circulation in a highly eccentric orbit}

\begin{figure}
    \centering
    \includegraphics[width=0.5\textwidth]{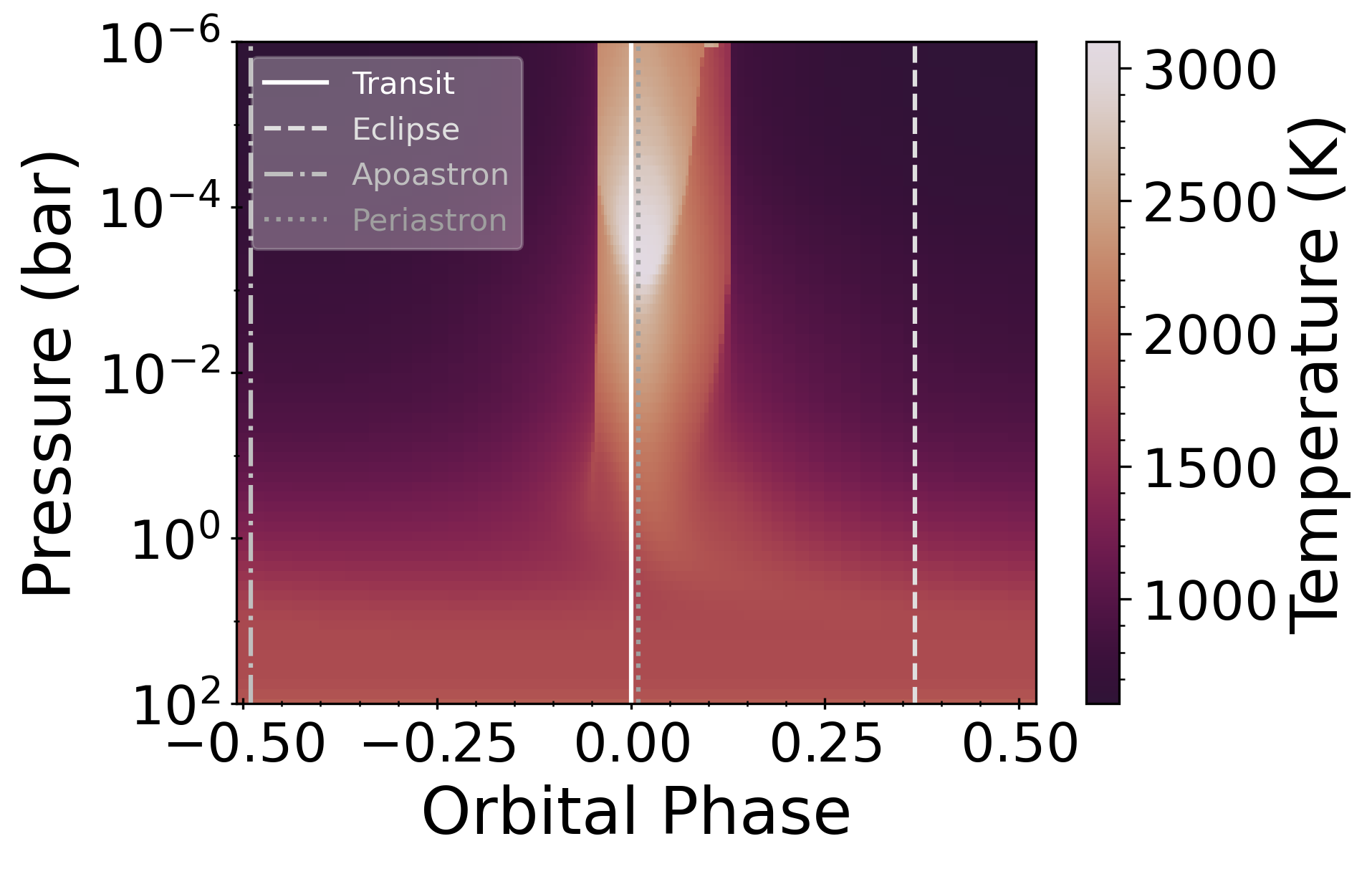}
    \includegraphics[width=0.5\textwidth]{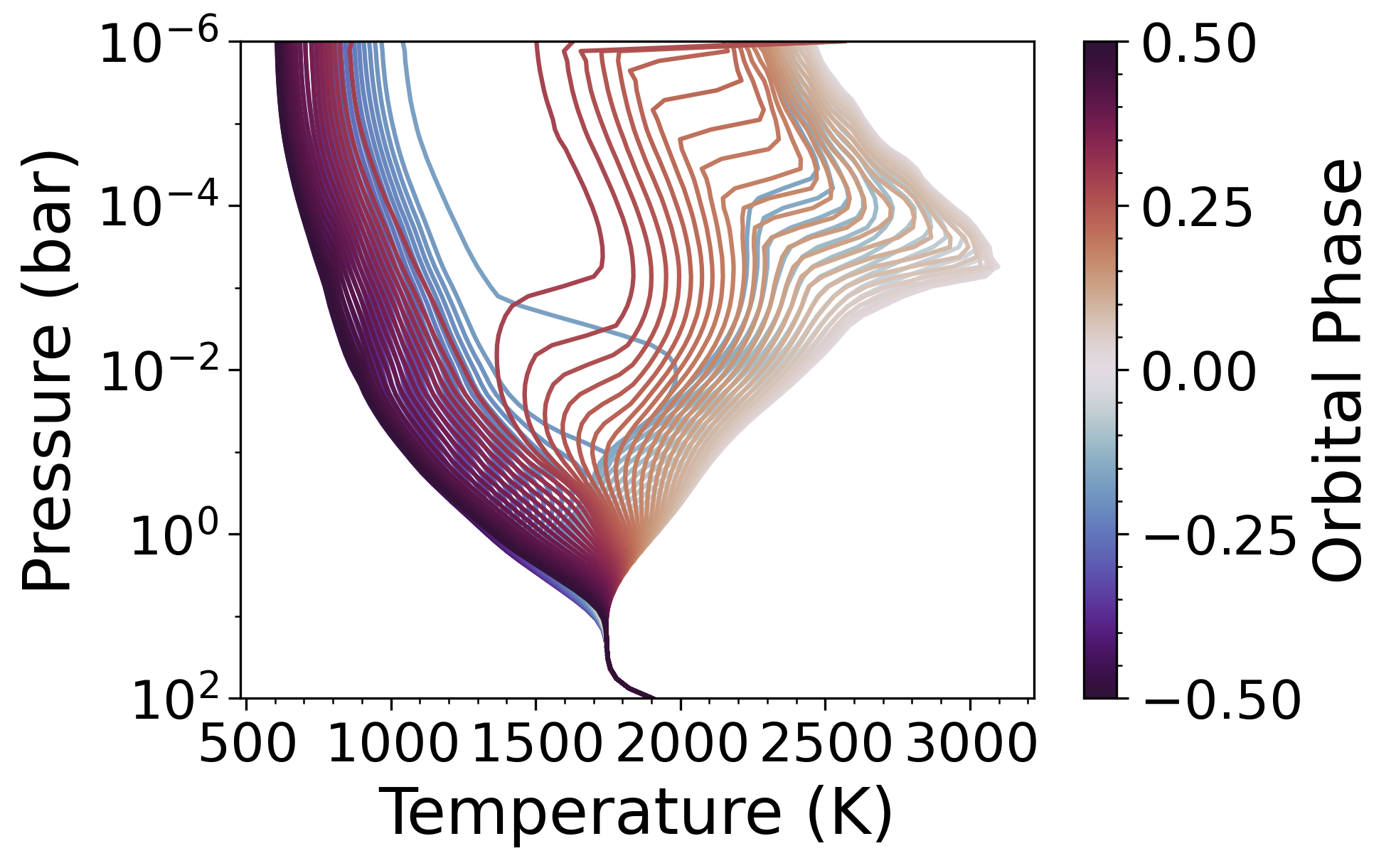}
    \caption{Temperature as a function of pressure over orbital phase from a 1D timestepping cloud-free atmospheric model generated using \texttt{EGP+}. Due to the presence of TiO and VO in the model atmosphere, at periastron the temperature of the upper atmosphere rises rapidly causing an inversion.}
    \label{fig:TP}
\end{figure}

\begin{figure*}
    \centering
    \includegraphics[width=\textwidth]{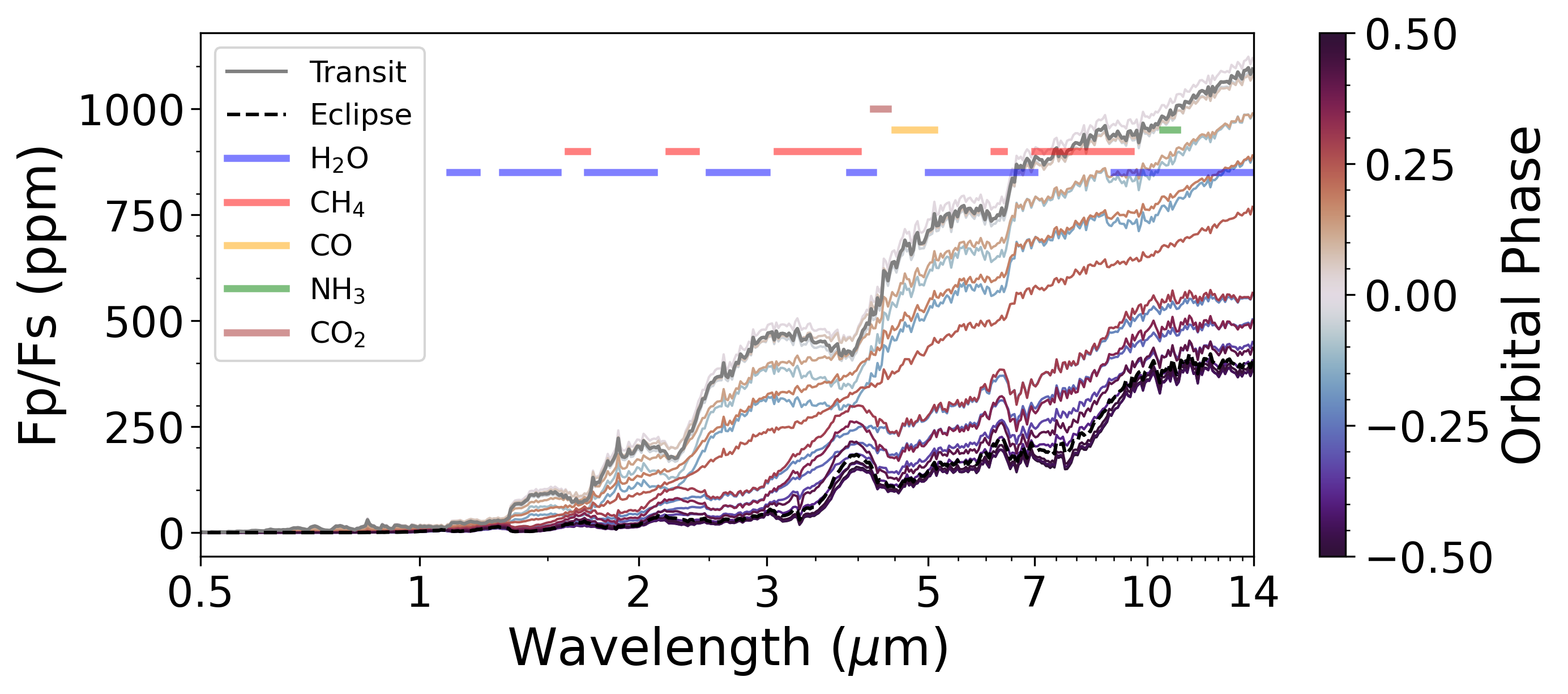}
    \caption{Planet-to-star flux ratio as a function of orbital phase from the same model as \autoref{fig:TP}. At apastron the spectra show many species in absorption (colored bars) and as the planet progresses towards periastron only the CO feature remains (in emission) with water providing the bulk of the increase in flux from a blackbody due to being in emission.}
    \label{fig:modspec}
\end{figure*}

The level of stellar irradiation received by \toib changes by more than an order of magnitude in less than 10 days. Unlike most circularized hot Jupiters, \toib is not expected to have a constant day/night side. 
To understand how this change in irradiation might affect such a planet we perform atmospheric modeling using \textit{EGP+}, a 1D time-stepping radiative-convective equilibrium model \citep{mayorga2021}. We follow the approach laid out in \citet{mayorga2021} to initialize the model and ensure spin up using the assumptions of Solar metallicity, C/O ratio, and a cloud-free atmosphere.
In \autoref{fig:TP}, we show how the model atmosphere's temperature varies as a function of pressure and orbital phase. Near periastron, the presence of TiO and VO cause a rapid change in the upper atmospheric temperatures leading to to a predicted inversion in the planet's TP profile. The planetary equilibrium temperature fluctuates from a maximum temperature of 2171\,K just after periastron passage at phase 0.01 to a minimum temperature of 1217\,K at phase -0.38.
In \autoref{fig:modspec}, we show how the model atmosphere's planet-to-star flux ratio varies as a function of orbital phase. Near apastron where the model atmosphere is relatively cool, many species have absorption features which shape the spectrum. As the planet transitions towards periastron, it passes through a blackbody phase before progressing to the near periastron state where signatures from water and carbon monoxide are the only features that remain due to the inverted nature and temperature of the TP profile.

Ultra-Hot Jupiters with similar equilibrium temperatures have neutral and ionized atomic species detected via high-resolution spectroscopic cross-correlations (e.g. KELT-20-b/MASCARA-2\,b, \citep{2020mascaraATM,2023mascaraATM,2024mascaraATM} and WASP-121\,b \citep{2020aHoeijmakers,2020bHoeijmakers,2022casasayas}). All of these previous planets reside in near-circular orbits, and receive near-constant stellar irradiation. The only existing measurements of rapid heating of exoplanet atmospheres were done in the infrared for HD 80606\,b ($e=0.93$) \citep{2024HD80606} and HAT-P-2\,b ($e=0.5$) \citep{2009laughlin, 2013lewis}) and both found extremely short atmospheric radiative time scales compared to Solar System planets. By comparing the strength of neutral and ionized species in the atmosphere of \toib, which undergoes these orbital induced thermal pulses, against those detected for planets in circular orbits, we can constrain the specific timescales of the chemical transitions and element transporting that take place in its highly irradiated atmospheres \citep{mayorga2021}. These time scales are crucial inputs for exoplanetary global circulation models \citep{2013lewis}, which would impact our deeper understanding of high signal-to-noise spectroscopic observations of exoplanetary atmospheres.

\section*{acknowledgments}

We respectfully acknowledge the traditional custodians of the lands on which we conducted this research and throughout Australia. We recognize their continued cultural and spiritual connection to the land, waterways, cosmos and community. We pay our deepest respects to all Elders, present and emerging people of the Giabal, Jarowair and Kambuwal nations, upon whose lands this research was conducted.
CH thanks the support of the ARC DECRA program DE200101840 and Future Fellowship program FT240100016.
GZ thanks the support of the ARC DECRA program DE210101893 and Future Fellowship program FT230100517.
Funding for the TESS mission is provided by NASA's Science Mission Directorate. We acknowledge the use of public TESS data from pipelines at the TESS Science Office and at the TESS Science Processing Operations Center. This research has made use of the Exoplanet Follow-up Observing Program website, which is operated by the California Institute of Technology, under contract with the National Aeronautics and Space Administration under the Exoplanet Exploration Program. Resources supporting this work were provided by the NASA High-End Computing (HEC) Program through the NASA Advanced Supercomputing (NAS) Division at Ames Research Center for the production of the SPOC data products. This paper includes data collected by the TESS mission that are publicly available from the Mikulski Archive for Space Telescopes (MAST). This work has made use of data from the European Space Agency (ESA) mission
{\it Gaia} (\url{https://www.cosmos.esa.int/gaia}), processed by the {\it Gaia}
Data Processing and Analysis Consortium (DPAC,
\url{https://www.cosmos.esa.int/web/gaia/dpac/consortium}). Funding for the DPAC has been provided by national institutions, in particular the institutions participating in the {\it Gaia} Multilateral Agreement. The Flatiron Institute is a division of the Simons foundation. This work makes use of observations from the LCOGT network. Part of the LCOGT telescope time was granted by NOIRLab through the Mid-Scale Innovations Program (MSIP). MSIP is funded by NSF. 
This work makes use of observations from the ASTEP telescope. ASTEP benefited from the support of the French and Italian polar agencies IPEV and PNRA in the framework of the Concordia station program and from OCA, INSU, Idex UCAJEDI (ANR- 15-IDEX-01) and ESA through the Science Faculty of the European Space Research and Technology Centre (ESTEC). This research also received funding from the European Research Council (ERC) under the European Union's Horizon 2020 research and innovation program (grant agreement No. 803193/BEBOP) and from the Science and Technology Facilities Council (STFC; grant No. ST/S00193X/1). A.J. acknowledges support from ANID -- Millennium  Science  Initiative -- ICN12\_009, AIM23-0001 and from FONDECYT project 1210718.

\vspace{5mm}
\facilities{MAST(TESS), CHIRON, FEROS, MINERVA-AUS, ASTEP, LCO, SOAR, ExoFOP}

\software{AstroImageJ \citep{aij}, Lightkurve \citep{lightkurve}, Tapir \citep{jensen2013}, Batman \citep{2015PASP..127.1161K}}

\bibliography{ref}{}
\bibliographystyle{aasjournal}

\end{document}